\documentclass{iau_FM}
\usepackage{graphicx}
\usepackage{natbib}

\newcommand\aj{{AJ}}%
\newcommand\araa{{ARAA}}%
\newcommand\apj{{ApJ}}%
\newcommand\apjl{{ApJ}}%
\newcommand\apjs{{ApJS}}%
\newcommand\aap{{A\&A}}%
\newcommand\mnras{{MNRAS}}%
\newcommand\prd{{Phys.~Rev.~D}}%
\newcommand\pasj{{PASJ}}%
\newcommand\nat{{Nature}}%
\newcommand\nar{{New Astron. Revs}}%

\title[FM 6.~~AGN host galaxies and structures] 
{Host galaxies and large-scale structures of active galactic nuclei}

\author[Hickox et al.]   
{Ryan C.\ Hickox$^1$, Stephanie M.\ LaMassa$^2$, John D.\ Silverman$^3$, Alexander Kolodzig$^4$}

\affiliation{$^1$Department of Physics and Astronomy, Dartmouth College, 6127 Wilder Laboratory, Hanover, NH 03755, USA \\ email: {\tt ryan.c.hickox@dartmouth.edu} \\
$^2$NASA Goddard Space Flight Center, Code 662, Greenbelt, MD 20771, USA\\
$^3$Kavli Institute for the Physics and Mathematics of the Universe (WPI), The University of Tokyo Institutes for Advanced Study, The University of Tokyo, Kashiwa, Chiba 277-8583, Japan \\
$^4$Kavli Institute for Astronomy and Astrophysics, Peking University, Beijing 100871, China}

\pubyear{2015}
\jname{Astronomy in Focus, Volume 1} 
\editors{Piero Benvenuti, ed.}
\begin{document}

\maketitle

\begin{abstract}

Our understanding of the cosmic evolution of supermassive black holes (SMBHs) has been revolutionized by the advent of large multiwavelength extragalactic surveys, which have enabled detailed statistical studies of the host galaxies and large-scale structures of active galactic nuclei (AGN). We give an overview of some recent results on SMBH evolution, including the connection between AGN activity and star formation in galaxies, the role of galaxy mergers in fueling AGN activity, the nature of luminous obscured AGN, and the connection between AGN and their host dark matter halos. We conclude by looking to the future of large-scale extragalactic X-ray and spectroscopic surveys.


\end{abstract}

\firstsection 
\section{Introduction}

The past two decades have seen great progress in understanding the growth and evolution of supermassive black holes (SMBHs) over cosmic time (for one of several recent reviews see  \citealt{alex12bh}). It is now well-established that SMBHs obtain the bulk of their mass through accretion of matter, observable as active galactic nuclei (AGN), and that there are connections between the cosmic growth of SMBHs and that of their host galaxies, due to common evolution histories or to feedback processes that link the growth of SMBHs to the state of gas and star formation in their host systems (see \citealt{fabi12feed} and \citealt{korm13bh} for recent reviews).

Recently, large multiwavelength extragalactic surveys have enabled breakthroughs in understanding the AGN-galaxy connection through detailed statistical studies of the host galaxies and large-scale structures of AGN. The resulting insights into SMBH evolution  are analogous to the understanding of galaxies that emerged from the large redshift surveys in the 2000's (e.g., \citealt{stra01galcol, zeha05a, blan06blue, coil06b}). Particularly valuable observational resources have been the {\em Chandra X-ray Observatory} \citep{tana2014chandra} and {\em XMM-Newton} \citep{jans01xmm}, for performing deep X-ray surveys detecting large numbers of AGN to high redshift; the {\em Herschel Space Observatory} \citep{pilb10herschel} to constrain star formation rates (SRFs) in AGN host galaxies; the {\em Spitzer Space Telescope} \citep{wern04spitzer} and {\em Wide-Field Infrared Survey Explorer} ({\em WISE}; \citealt{wrig10wise}) for identifying large numbers of luminous, obscured quasars; the {\em Nuclear Spectroscopic Telescopic Array} ({\em NuSTAR} \citealt{harr13nustar}) for studying the high-energy emission from heavily obscured AGN; and extensive follow-up of AGN with ground-based multi-object spectrographs. 

In this Proceedings, we begin with an overview of recent progress on the connection between AGN activity and star formation (SF) in galaxies. We then discuss the link between AGN activity and galaxy mergers, highlight observational studies of the luminous, obscured AGN that may represent an important phase in the evolution of massive galaxies, and discuss the utility of AGN clustering measurements in understanding the connection between AGN and their host dark matter (DM) halos. Finally, we look toward the future of statistical studies of AGN host galaxies and structures with the next generation of very large X-ray and spectroscopic surveys.


\section{The AGN star-formation connection}

There is compelling indirect evidence for a global connection between AGN activity and SF in galaxies, from the tight correlation between SMBH masses and galaxy properties \citep[e.g.,][]{mcco13bhgal, korm13bh} and the similar cosmic evolutionary histories of these two processes \citep[e.g.,][]{merl13agn, korm13bh}. We can now probe this connection {\em directly} by observing the SF properties of AGN host galaxies. One useful observational tool is the distribution of galaxy colors and luminosities   \citep[e.g.,][]{stra01galcol, blan06blue}, which clearly separates galaxies into two populations: blue, star-forming, relatively low-mass galaxies, and red, passive, higher-mass systems. When we locate AGN host galaxies (out to redshifts $z\sim1$) in color-luminosity space, we find that the hosts of radiatively efficient, rapidly growing SMBHs (identified as AGN based on X-ray, infrared, or optical line emission) are predominantly located among the star-forming systems \citep[e.g.,][]{nand07host, hick09corr, goul14agn, mend15agnclust}. There is a preference for more luminous AGN to be found in the more massive, redder end of the star-forming galaxy population \citep[e.g.,][]{scha09agn, scha10morph}. perhaps connected to the higher Eddington limit associated with their more massive SMBHs \citep[e.g.,][]{aird12agn, trum15sdss}.

\begin{figure}[t]
\begin{center}
 \includegraphics[height=2.1in]{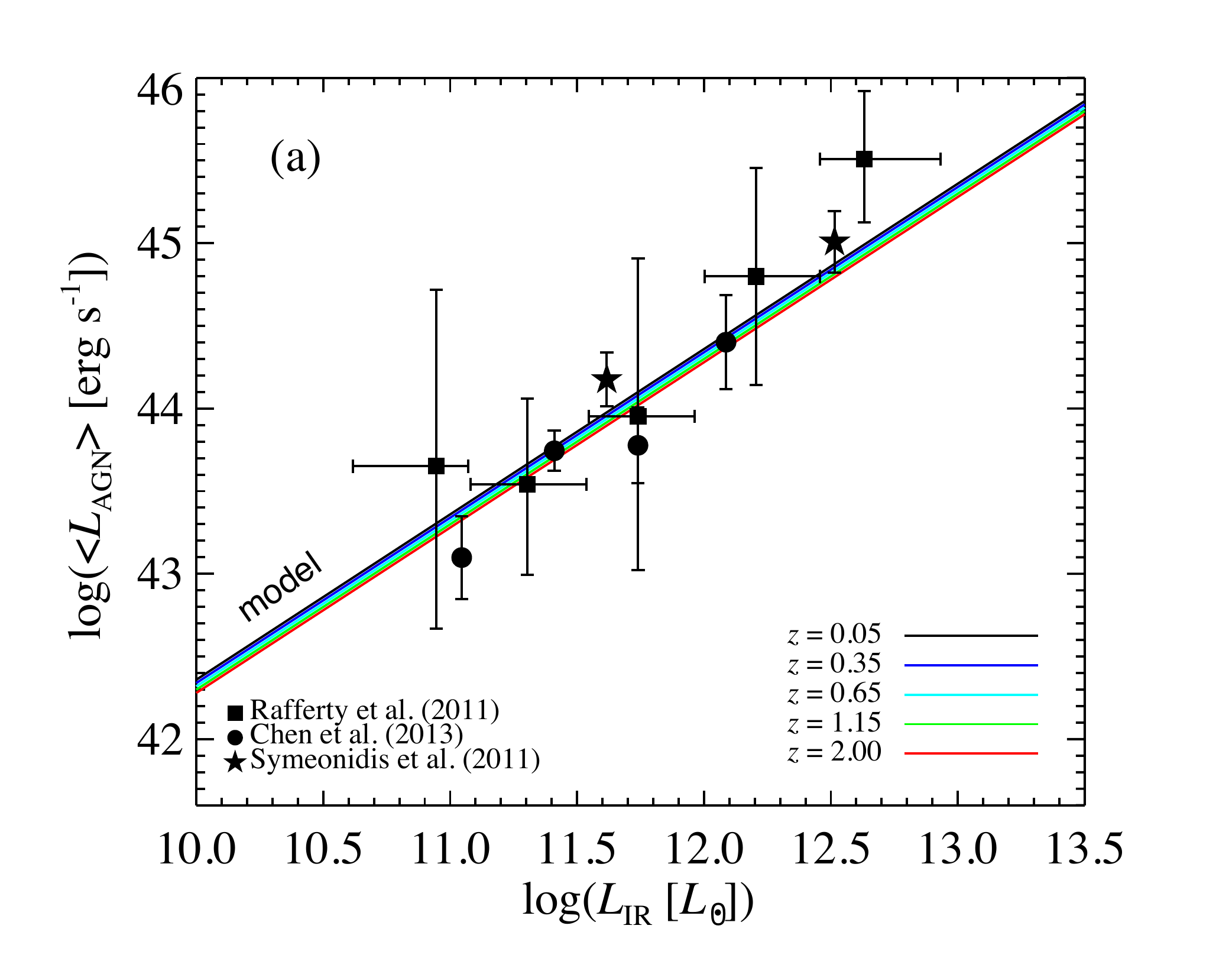}  \includegraphics[height=1.95in]{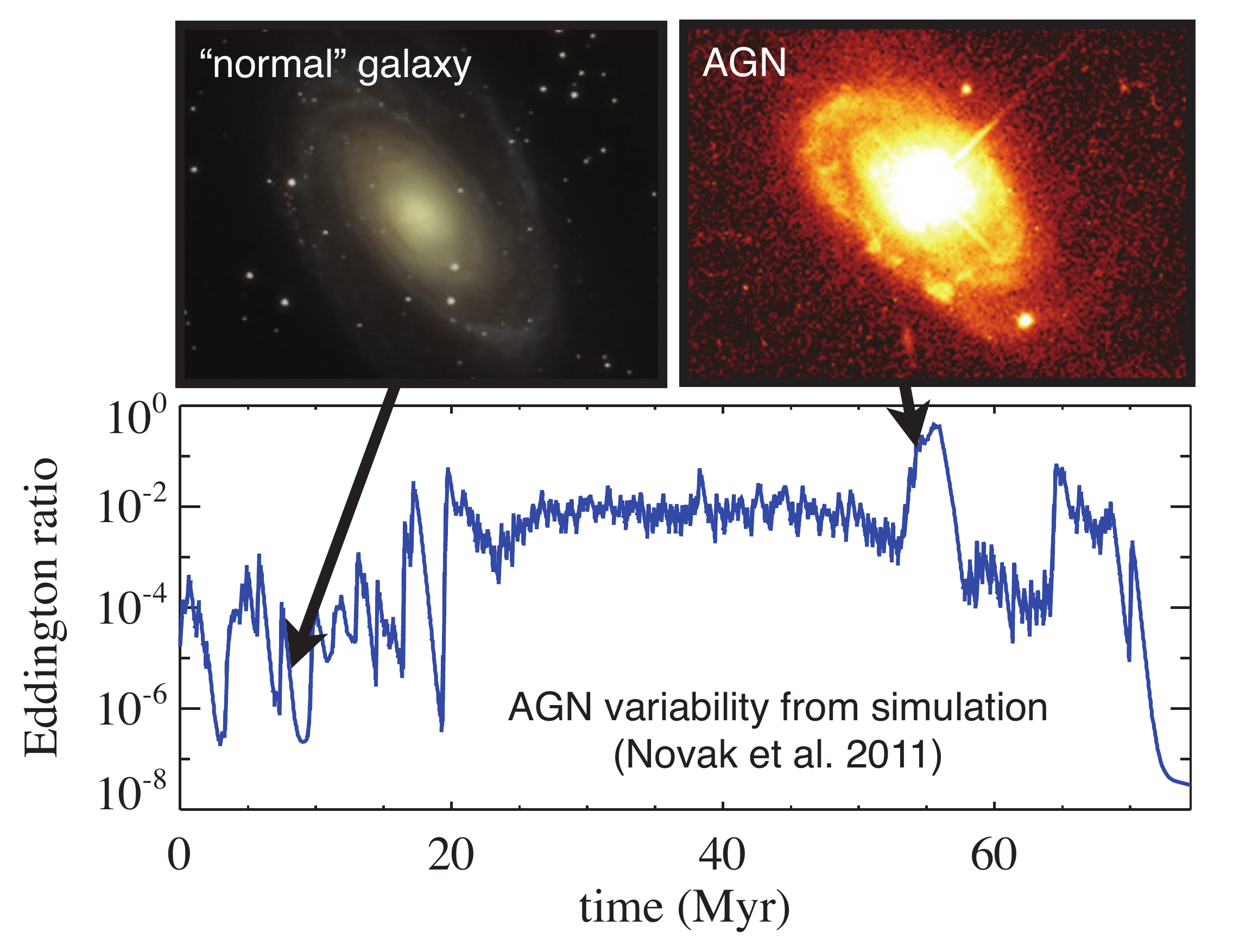} 
 \caption{{\em Left:} The observed correlation between SFR (as measured by far-IR luminosity) in star-forming galaxies and the average AGN luminosity (see \citealt{hick14agnsf} for references). Colored lines show the linear relationship assumed in the model of \citep{hick14agnsf}. {\em Right:} Illustration of AGN variability. The bottom panel shows the Eddington ratio versus time for the simulation of \citet{nova11bhsim}.  Image credits, from left: M81: Wikimedia Commons; PG 0052+251: J.\ Bahcall (IAS, Princeton), M.\ Disney (Univ.\ of Wales), NASA/ESA. }
   \label{fig:sf}
\end{center}
\end{figure}

We can further study the host galaxies of {\em mechanically}-dominated AGN identified by radio synchrotron emission from relativistic jets. In contrast to the radiatively-efficient AGN, radio AGN are found predominantly in massive, passive galaxies, generally avoiding the star-forming systems \citep[e.g.,][]{hick09corr, goul14agn, mend15agnclust}. (Note that AGN with radio jets {\em and} high-excitation emission lines are generally found among the star-forming galaxies; \citealt{smol09radio}.) These results point toward a general picture in which rapidly-growing, radiatively-efficient AGN are found in star-forming galaxies, fueled by the supply of cold gas that also produces the SF, while mechanically-dominated, slowing growing AGN are found in passive galaxies, producing the feedback required to stop cooling of hot gas in their massive halos \citep[e.g.,][]{bowe06gal, crot06}.

Recent studies have built on this work by taking advantage of {\em Herschel} for reliable measures of SFR in AGN host galaxies based on far-IR emission, where there is little contamination from the AGN \citep[e.g.,][]{mull11agnsed}. These results have shown that when averaging over the full star-forming galaxy population, the average SMBH growth is correlated with SFR \citep[e.g.,][]{mull12agnms}, with a linear relationship between SFR and average accretion rate \citep[e.g.,][Figure~\ref{fig:sf}]{chen13agnsf}. However, the picture becomes more complicated when looking at {\em individual} AGN. The colors, SFRs, spatial clustering, and merger rates of AGN hosts are essentially indistinguishable from typical star-forming galaxies of similar mass \citep[e.g.,][]{card10xhost, xue10xhost, mull11agnsf, cist11agn, trei12merge, mend15agnclust}. The average SFR of AGN host galaxies at depends strongly on redshift similarly to inactive galaxies, but shows little if any dependence on AGN luminosity \citep[e.g.,][]{shao10agnsf, rosa12agnsf, stan15agnsf}.

We therefore are presented with a puzzle in which there is a strong global correlation between SF and AGN activity, but the links are weak if non-existent for individual sources. This raises the question: "why are only a fraction of star-forming galaxies observed as AGN?" The answer appears to lie in the {\em stochastic variability} of AGN. 
Simulations \citep[e.g.,][]{nova11bhsim, gabo13bhsim} and observations of AGN light echoes \citep[e.g.][]{scha10voor, keel15fade} suggest that AGN accretion can vary over many orders of magnitude on timescales of 1 Myr or less, much shorter than the typical evolution timescale for galaxies. We might therefore think of {\em all} star-forming galaxies as hosting an AGN, when averaged over $>100$ Myr timescales. \citet{hick14agnsf} presented a simple analytic model for the AGN population based on a direct connection between SMBH accretion and star formation and including stochastic variability. This simple model is able to reproduce the observed relationships between AGN luminosity, SFR, and merger rates, as well as the general evolution of the AGN luminosity function. Other recent simulations have found similar results \citep[e.g.,][]{thac14agnsf,volo15agnsf}, highlighting the need for future studies to not simply compare AGN host galaxies with their inactive counterparts, but to measure the {\em distribution} of AGN accretion rates as a function of host galaxy properties.  

Given the clear connection between SMBH growth and SF, an important question arises regarding the nature of AGN {\em feedback}, and whether in some cases the AGN can shut down star formation by heating or removing the gas supply (see \citealt{fabi12feed} for a review). Some simulations show that  outflows from radiatively efficient AGN can have a strong effect on galaxy-scale gas \citep[e.g.,][]{dima05qso, boot11bhscale}, but other models of AGN in disk galaxies show a limited impact of the AGN on the star-forming disk \citep{gabo14feed}. Some theoretical studies also indicate that AGN activity can {\em trigger}
SF through positive feedback \citep[e.g.,][]{zubo13agnsb, naya13feed}. Observationally, AGN radiation can ionize gas in the galaxy out to large scales \citep[e.g.,][]{hain13salt, hain14nlr}, and outflows clearly drive out large quantities of ionized and molecular gas in some systems \citep[e.g.,][]{harr14agnoutflow, feru15mrk231}. There are systems in which powerful outflows appear to be driven by compact starbursts rather than AGN \citep[e.g.,][]{dima12hizea, geac14outflow}. Together, these results point toward a complex relationship between AGN activity and star formation, and understanding the details of this connection is the focus of a great deal of ongoing research.

\section{Fueling of AGN by galaxy mergers}

\begin{figure}[t]
\begin{center}
 \includegraphics[height=1.95in]{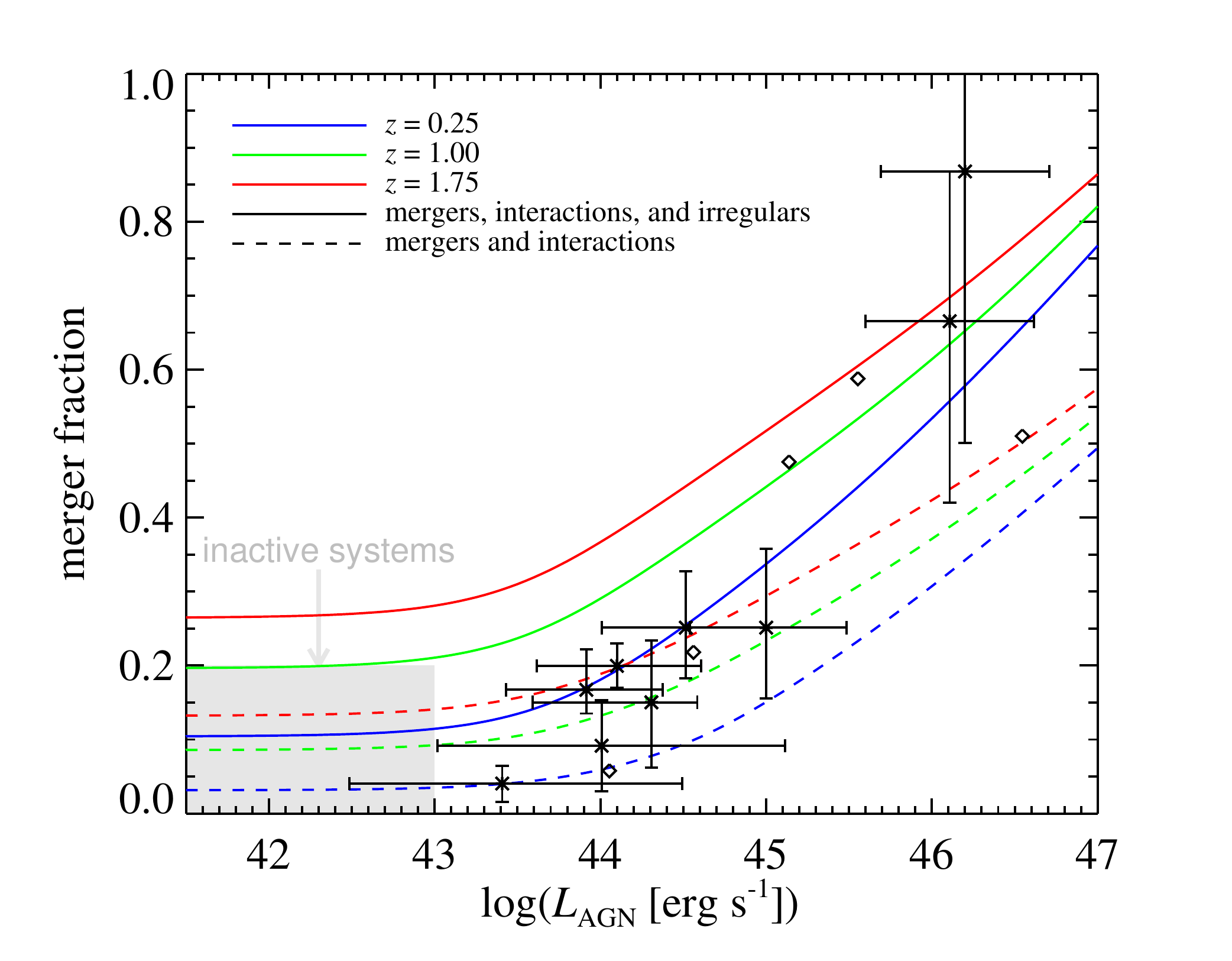} 
 \includegraphics[height=2.in]{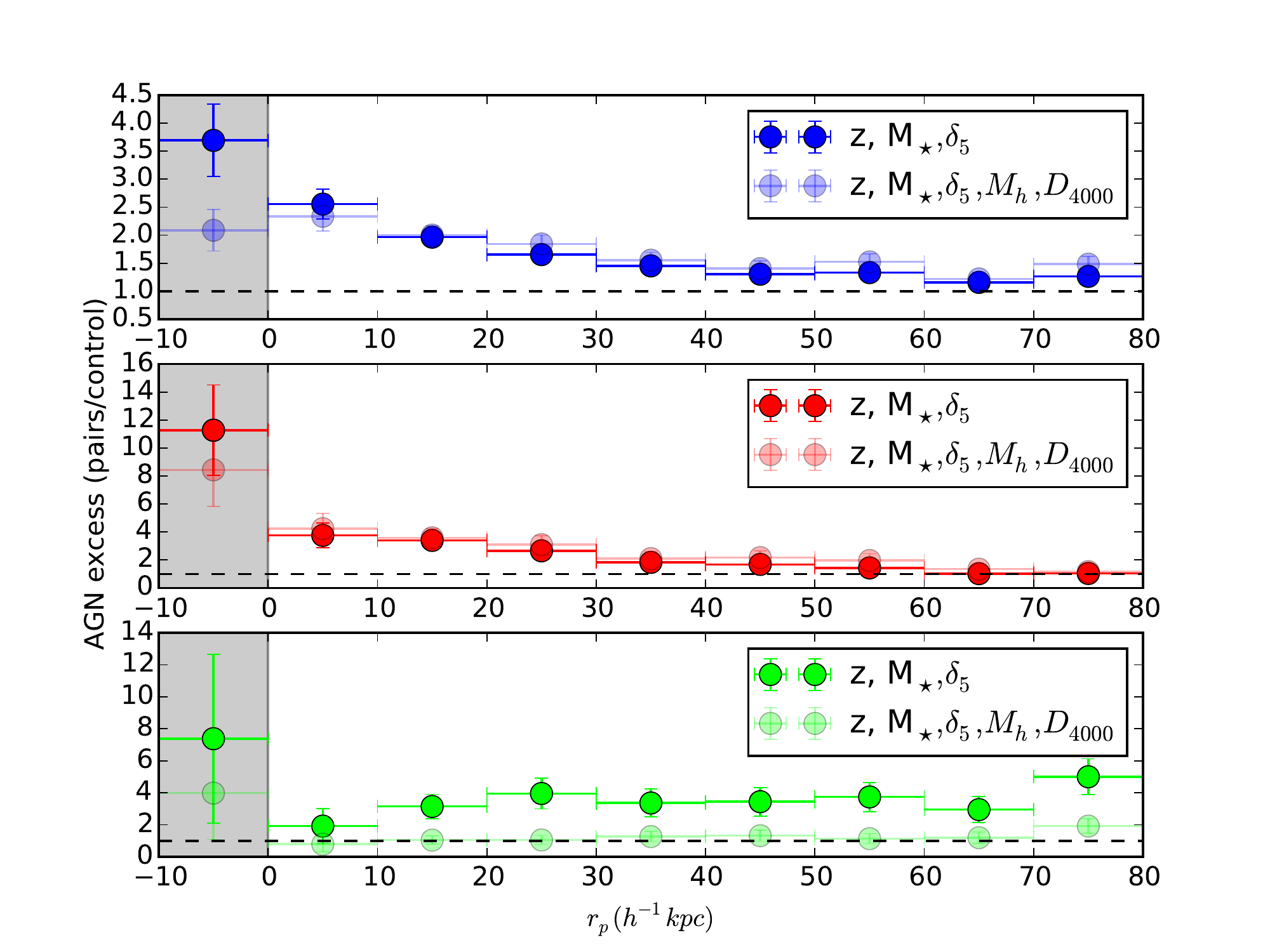} 
 \caption{{\em Left:} Relationship between the fraction of AGN in mergers and AGN luminosity taken from
  the compilation of \citet{trei12merge}. The gray shaded area
  indicates the typical range of merger fractions for inactive
  galaxies in the control samples studied by \citet{cist11agn} and
  \citet{koce12xmorph}. The colored curves show the predictions of the \citet{hick14agnsf}
  model assuming a correlation between merger fraction and $L_{\rm
    IR}$ determined by \citet{kart12irmorph}. {\em Right:} The fraction of AGN in galaxy pairs and post-mergers (shown in the grey area at left)
  relative to matched
  control galaxies versus projected separation.  Optically selected,
  mid-IR selected and radio selected are shown in the top, middle and bottom
  panels respectively. All three samples show an elevated AGN fraction over the control galaxies; the optical and mid-IR AGN shown an increasing enhancement in AGN activity with decreasing pair separation.}
   \label{fig:merger}
\end{center}
\end{figure}

An important aspect of the connection between SMBHs and galaxies has been the role of galaxy mergers in fueling AGN activity. This has long been a matter of debate, with some studies showing a clear link between AGN activity and mergers \citep[e.g.,][]{bahc97qsohost, urru08qsohost, koss10batagn, glik15qsomerge}, while other studies show effectively no difference in the merger rates between AGN and inactive galaxies \citep[e.g.,][]{grog05agnhost, gabo09xmorph, cist11agn, koce12xmorph, mech15qsomerge}. \citet{trei12merge} suggested that these differences can be reconciled when considering AGN as a function of luminosity, with powerful quasars commonly found in mergers, while less luminous AGN have merger rates of 20\% or less, similar to "normal" galaxies (Figure~\ref{fig:merger}, {\em left}). While this work necessarily relies on heterogeneous definitions of "mergers", the results are broadly suggestive of a trend with AGN luminosity.

A powerful way to test the merger-AGN connection is by directly tracing AGN activity along the merger sequence, using kinematic pairs of galaxies. Kinematic pairs are well-known to show an enhancement of SFR that increases with decreasing separation \citep[e.g.,][]{elli08pair, wood10galpair, kamp13galmerge}. The incidence of AGN activity shows very similar behavior (Figure~\ref{fig:merger}, {\em right}), for AGN identified from optical emission lines \citep{elli15agninter}, X-ray luminosity \citep{silv11agninter}, or mid-infrared (IR) colors \citep{saty14wisepairs}. Radio-loud AGN with low-excitation optical spectra are also more common in galaxy pairs, however in constrast to the radiatively-efficient AGN, this enhancement does not depend on separation \citep{elli15agninter}, suggesting that fueling is not directly related to galaxy interactions. To probe the end stages of the merger sequence, one can search for galaxies with double nuclei through careful analysis of optical images. These double nuclei show a clear enhancement in the frequency of X-ray AGN \citep{lack14merge}, further supporting a connection between AGN and mergers.

Despite these clear connections between AGN and mergers, a statistical analysis of the AGN population suggests that mergers are associated with only only a minority ($\sim$20\%) of the total AGN activity \citep{lack14merge}. In general, SMBH growth may simply trace the availability of cold gas, so that most AGN are found in normal star-forming galaxies, but the probability of finding an AGN is enhanced in mergers just as is observed for SF. Thus it may be possible to reconcile apparently contradictory observations with a picture in which SMBH growth is connected to galaxy interactions, but due to stochastic variability, an excess in the fraction of mergers is only detectable for the most luminous AGN \citep[Figure~\ref{fig:merger}, {\em right}]{hick14agnsf}.

\section{Obscured AGN and the evolutionary sequence}

In studying the population of luminous AGN fueled by major mergers, it is common to invoke an {\em evolutionary sequence} in which the merger produces a powerful, dust obscured starburst, followed by a period of powerful obscured AGN activity and finally by an unobscured quasar \citep[e.g.,][]{sand88, hopk08frame1}. In this scenario, obscured quasars represent an important phase in the life of massive galaxies. With sensitive mid-IR observations from {\em Spitzer} and {\em WISE}, we have now identified large numbers of luminous obscured quasars with little or no rest-frame optical emission from the nucleus \citep[e.g.,][]{hick07abs, ster12wise, asse13wiseagn, asse15wiseqso}.  X-ray observations (particularly with {\em NuSTAR}) suggest these obscured quasars are heavily buried or even Compton-thick \citep[e.g.,][]{ster14nustarwise, lans14nustarqso, lans15nustarqso}, and {\em Herschel} observations suggest that these sources are associated with enhanced star formation \citep{chen15qsosf}. 

\begin{figure}[t]
\begin{center}
 \includegraphics[width=0.75\textwidth]{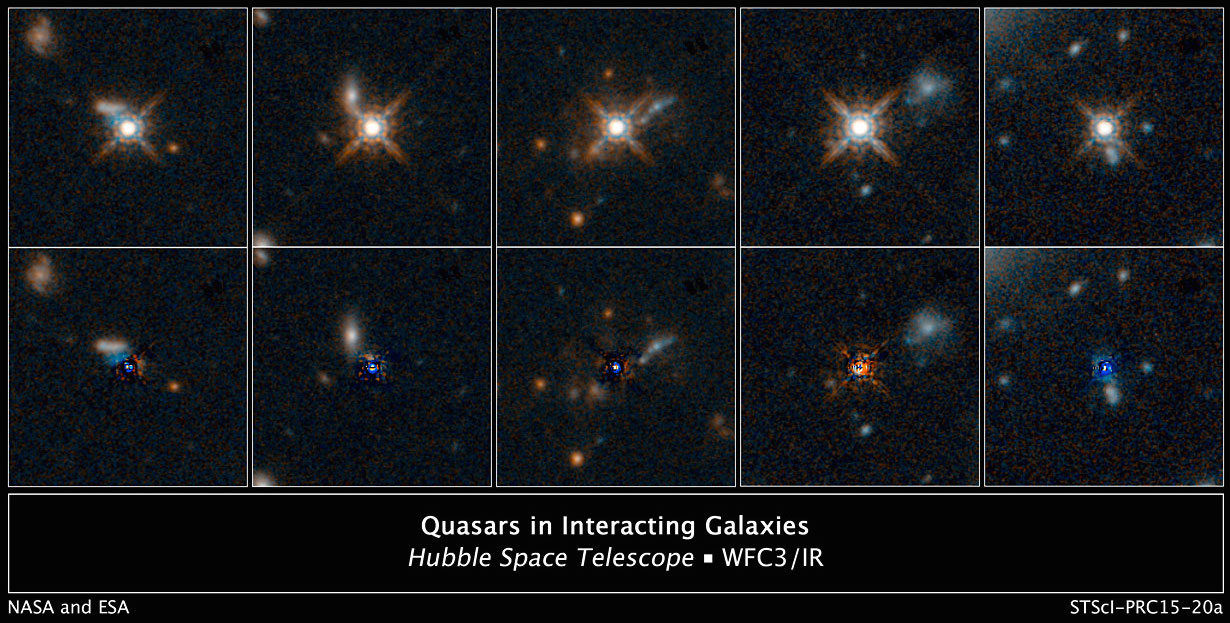} 
 \caption{{\em Hubble Space Telescope} WFC3 images of red QSOs, described in \citet{glik15qsomerge}, showing the observed images (top) and the residuals after subtracting smooth quasar and galaxy models (bottom). The majority of the red QSOs show  disturbances characteristic of major mergers. (Image credit: E.\ Glikman/NASA)   \label{fig:redqso}}
\end{center}
\end{figure}

In addition to these heavily buried quasars, it is particularly interesting to study quasars that are {\em emerging} from the dust. These show intermediate levels of dust extinction, and so are detectable as luminous broad-line AGN but with significant reddening of their optical and UV emission. Samples of these "red QSOs" have been identified using near-IR observations in concert with the radio \citep[e.g.,][]{glik07redqso, glik12redqso, glik13redqso}, mid-IR \citep[e.g.,][]{bane12redqso, bane13redqso, bane15redqso}, and X-rays \citep[e.g.,][]{brus05xqso, brus10cosmosagn}. These sources tend to be among the most luminous quasars known \citep{glik12redqso, bane15redqso}, are characteristically found in ongoing major mergers \citep[e.g.,][Figure~\ref{fig:redqso}]{urru08qsohost, glik15qsomerge}, and often show powerful outflows \citep{brus15agnoutflow, brus15moloutflow, pern15iroutflow, pern15z2outflow}. These characteristics are consistent with their identification as a population in transition between a deeply buried, luminous AGN and an unobscured quasar phase. 

Because this transitional phase is short-lived and therefore rare, identifying a large sample of reddened QSOs requires the large volumes probed by wide-area surveys. X-ray observations are particularly powerful for detecting these AGN, and an excellent resource is the X-ray data set in the SDSS Stripe 82 region \citep{lama13stripe82, lama15stripe82x}. The X-ray surveys in Stripe 82 combine archival data with dedicated {\em XMM-Newton} pointings, and contain over 6000 sources over an area of more than 31 deg$^2$. The vast majority of these sources have multiwavelength counterparts, and $\approx$30\% have spectroscopic redshifts. Using this large data set, samples of candidate red QSOs have been identified based on optical, near-IR, and mid-IR properties \citep[e.g.,][]{lama15xobs}. Near-IR spectroscopic surveys of these targets are currently underway (LaMassa \etal\ 2016, in prep), and will provide a valuable data set for studying this interesting phase in the evolution of massive galaxies.

\section{AGN clustering and dark matter halos}

\begin{figure}[t]
\begin{center}
 \includegraphics[width=0.7\textwidth]{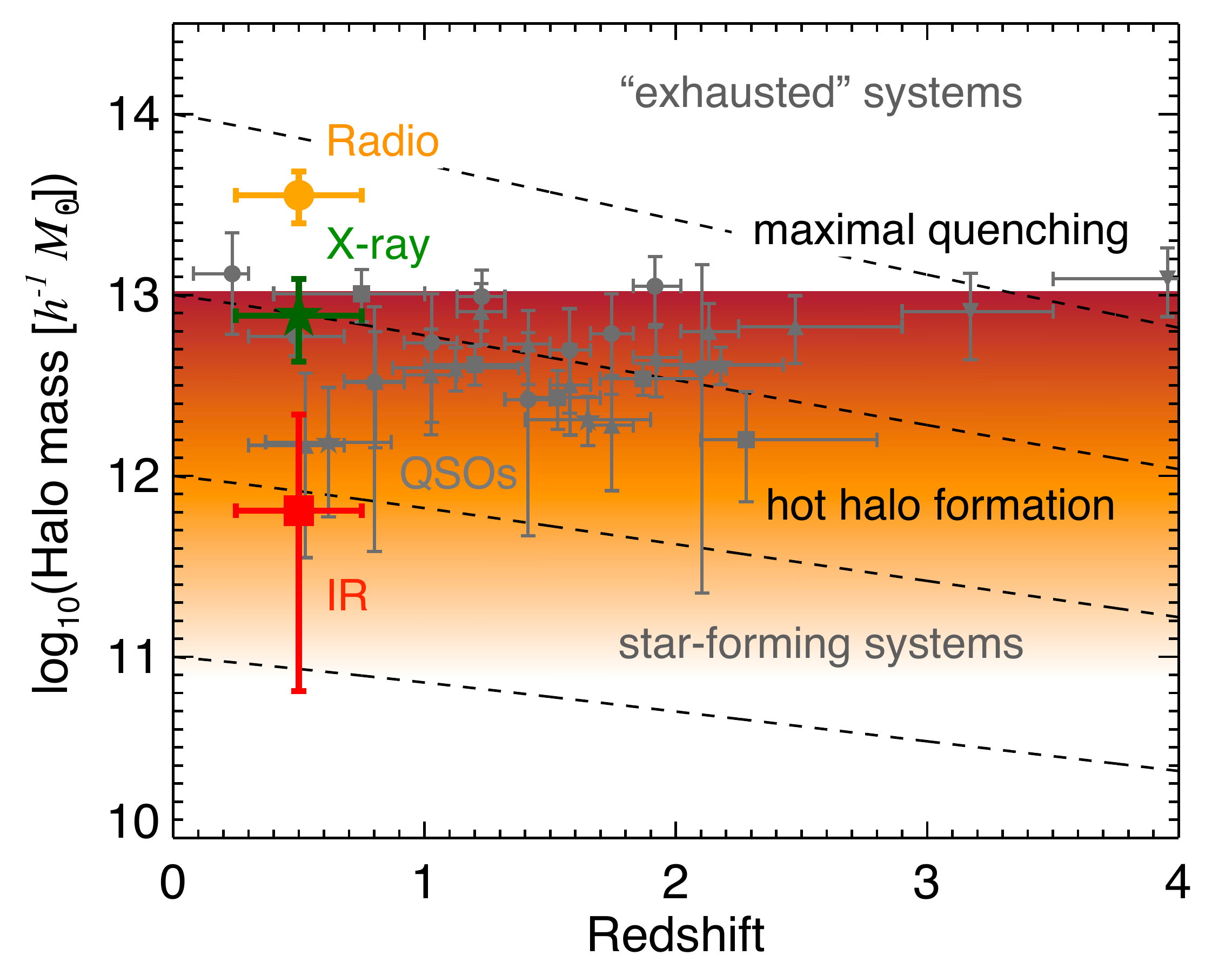} 
 \caption{Illustration of the evolution of dark matter halo mass versus
  redshift for AGN populations \citep{alex12bh}. Lines show the evolution of halo mass
  with redshift for individual halos. Highlighted is
  the region of maximum ``quenching'', in which halos transition from
  having large reservoirs of cold gas to being dominated by virialized
  hot atmospheres. The gray points show
  halo masses of optically-bright quasars derived from autocorrelation
  measurements (see \citealt{alex12bh} for references). The colored points show the halo masses for radio, X-ray,
  and infrared-selected AGN at $z\sim 0.5$ \citep{hick09corr}.}
   \label{fig:halo}
\end{center}
\end{figure}

Another valuable technique for studying the evolution of SMBHs is to connect AGN populations to their parent dark matter halos via measurements of spatial clustering (see \citealt{capp12xclust} for a comprehensive review of X-ray AGN clustering). The growth of DM halos is well understood from simulations and analytic theory \citep[e.g.,][]{shet01halo, tink08halo}, so knowledge of how AGN populate DM halos provides a powerful constraint on models of galaxy formation \citep[e.g.,][]{fani13xclust}. Measurements of the linear clustering amplitude of optical quasars have shown that they are found in halos of constant mass $10^{12}$--$10^{13}$ $M_\odot$ at all redshifts \citep[e.g.,][Figure~\ref{fig:halo}]{croo05, myer07clust1}. Halos of this mass have the highest ratios of stellar mass to dark matter mass \citep[e.g.,][]{most10mass} and are also the sites of powerful, dust-obscured high-redshift starbursts (submillimeter galaxies; \citealt{hick12smg}). This suggests that the maximum BH growth occurs in the same systems in which SF is the most efficient. In contrast, slowly-growing, mechanically-dominated AGN are found in halos of mass $>10^{13}$ $M_\odot$ \citep[e.g.,][Figure~\ref{fig:halo}]{hick09corr, mend15agnclust}, for which virial temperatures of the intergalactic gas are higher, star formation rates are lower, and mechanical energy input is required to offset cooling in the centers of the halos that would produce further star formation \citep[e.g.,][]{bowe06gal, crot06}.

Spatial clustering also provides an independent test of quasar evolutionary models by comparing the halo masses of obscured and unobscured sources. In the simplest "unified scenario" in which AGN obscuration is due only to orientation \citep[e.g.,][]{netz15unified}, one would expect no difference in halo mass. However in evolutionary scenarios in which obscured quasars evolve into unobscured sources, the halo masses of the two types of quasars can differ. The clustering of obscured quasars identified with {\em Spitzer} and {\em WISE} suggests that they reside in higher mass halos than similar unobscured quasars \citep[e.g.,][]{hick11qsoclust, dono14qsoclust, dipo14qsoclust, dipo15wise}. However, measurements of X-ray selected AGN indicate higher clustering for {\em unobscured} sources \citep{alle14xclust}, and other studies of lower-luminosity IR-selected AGN show no such dependence \citep{mend15agnclust}.   A new, independent tool for studying quasar clustering comes from the cross-correlation of quasar positions with lensing maps derived from the cosmic microwave background \citep[e.g.,][]{sher12qsocmb, geac13qsocmb}; these analyses appear to confirm higher host halo masses for obscured quasars \citep{dipo15qsocmb, dipo15wise}. However, these lensing and clustering results still have relatively large uncertainties, motivating higher-precision measurements in the future. The relationship between clustering and obscuration thus remains an interesting open question.

Recent studies have begun to explore not only the mass of the DM halos that host AGN, but how AGN are distributed {\em within} those halos. This work uses the halo occupation distribution (HOD) formalism, which has proven successful for studying the clustering of galaxies \citep[e.g.,][]{berl02hod, zeha11galclust}. Pioneering studies of the AGN HOD confirm that AGN are found in halos with average masses $10^{12}$--$10^{13}$ $M_\odot$, but suggest that the host halos can span a wide range in mass \citep[e.g.,][]{miya11xhod, krum12xclust, rich13agnhod, shen13qsohod}. HOD analyses of clustering \citep{star11xclust, rich12qsohod, rich13agnhod,shen13qsohod} and direct measurements of AGN occupation in groups \citep{alle12agngroup, silv14agngroup} also indicate that {\em central} galaxies are more likely to host an AGN, although some clustering studies are consistent with a large number of satellites \citep[e.g.,][]{miya11xhod, krum15xclust}. There are further some hints that the number of AGN in satellites rises slowly or even decreases at large halo mass \citep[e.g.,][]{khab14agnenv, krum15xclust}, in contrast with the behavior of inactive galaxies, which may provide an important clue to the process of AGN fueling.  While AGN HOD studies are currently challenging due to small size of AGN samples and thus limited statistical power, they represent the first step in understanding the complete connection between AGN activity and host DM structures, and provide strong motivation for future generations of large extragalactic surveys that can identify and characterize large numbers of AGN.

\section{The future of large AGN surveys}

\begin{figure}[t]
\begin{center}
 \includegraphics[height=1.8in]{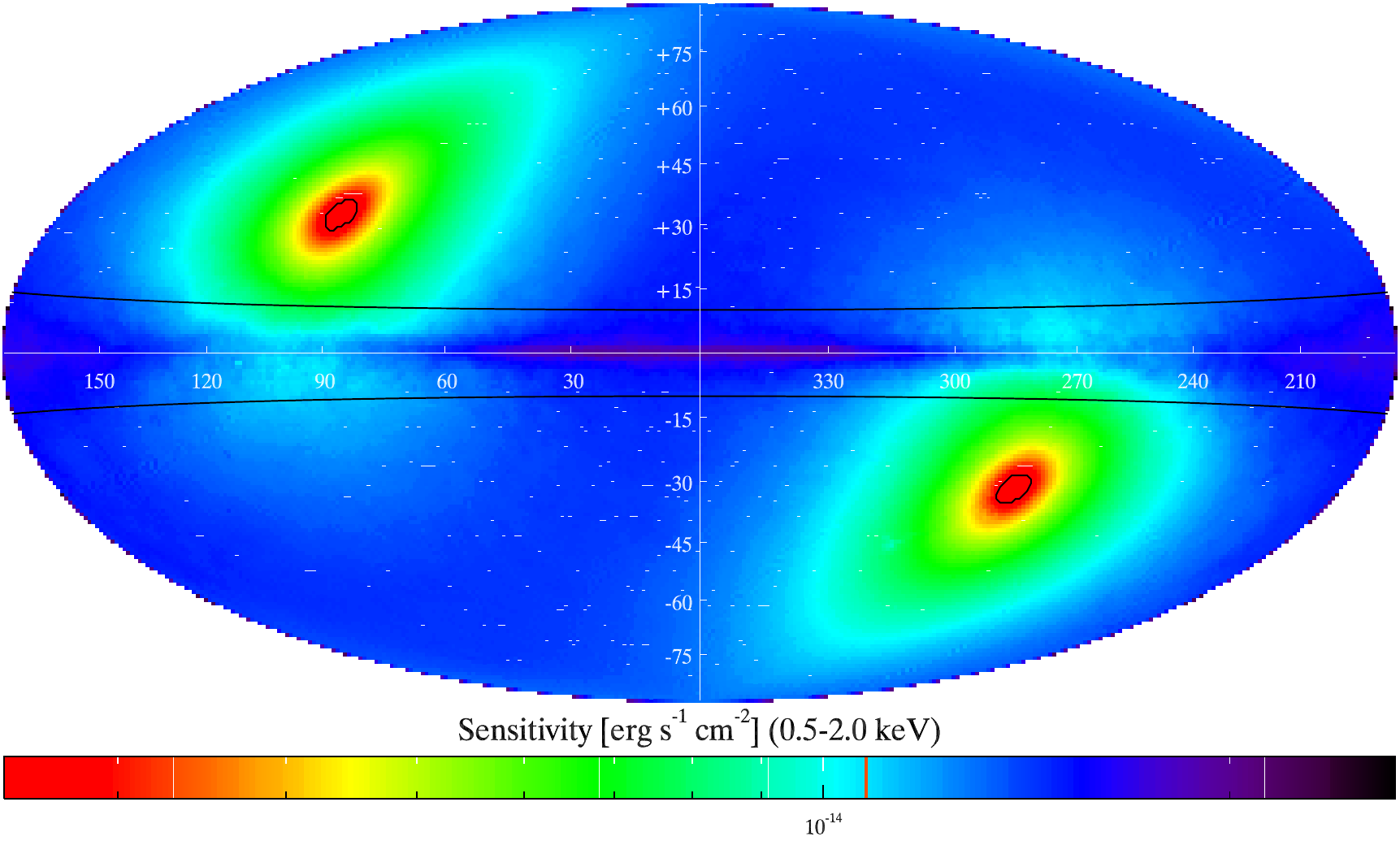} 
  \includegraphics[height=1.8in]{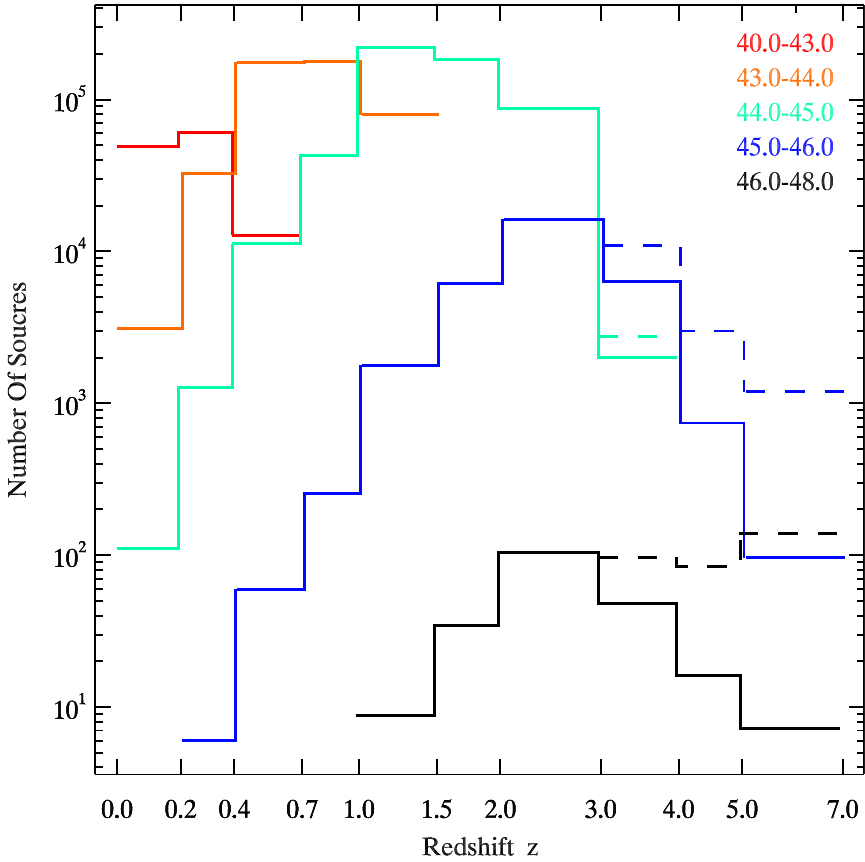} 

 \caption{{\em Left:} Estimated four-year soft (0.5--2 keV) band sensitivity eRASS in Galactic coordinates \citep{kolo13erass}. {\em Right:} Predicted numbers of AGN detected in the soft band by eRASS as a function of redshift, in five bins of X-ray luminosity, for an area of 14,000 deg$^2$, or approximately the coverage of SDSS \citep{kolo13erass}.}
   \label{fig:erass}
\end{center}
\end{figure}

In the coming years, a number of large-scale surveys over a wide area will dramatically expand the samples of AGN. Wide-area X-ray surveys can efficiently and unambiguously identify AGN with limited contamination from host galaxies, and even relatively short exposures with current observatories can produce large AGN samples comprising thousands of sources, as evidenced by the the wide-area {\em Chandra} XBo\"{o}tes \citep{murr05, kent05}, and {\em XMM}--XXL \citep{pier14xxl}, and {\em XMM} Stripe 82X \citep{lama15stripe82x} surveys. The next generation wide-area X-ray survey will be the eRosita All Sky Survey (eRASS), to be carried out by the eROSITA instrument on the Spectrum-Roentgen-Gamma spacecraft (\citealt{pred07erosita, merl12erosita}; Figure~\ref{fig:erass}). eRASS is expected to detect $\sim$$3\times10^6$ AGN over 34,000 deg$^2$ with a median redshift $z\sim 1$ and $>10^4$ AGN at $z>3$, increasing the known samples of X-ray AGN by more than an order of magnitude \citep{kolo13erass}. This sample will enable precision studies of the AGN luminosity function and clustering, using photometric redshifts from wide-field imaging surveys and  spectroscopic redshifts from the SPIDERS project (part of SDSS-III \citealt{eise11sdss3}, covering $>$50,000 eRASS AGN) and subsequently from the 4 metre Multi-Object Spectroscopic Telescope (4MOST; \citealt{dejo144most}, covering $\sim$700,000 eRASS AGN). These samples are large enough to enable high-precision (signal-to-noise ratio $>$10) measurements of AGN clustering amplitude in small bins of redshift or luminosity. With sufficiently large numbers of accurate redshifts, eRASS will also enable the first measurement with X-ray AGN of baryon acoustic oscillations \citep{kolo13erassclust, huts14xclust}.

Beyond eROSITA, SDSS-III, and 4MOST, future wide-field survey instruments include the Athena X-ray mission \citep{barc15athena}, the Euclid \citep{laur12euclid} and WFIRST \citep{cont13wfirst} optical and near-IR satellite missions, as well as the Subaru Prime Focus Spectrograph, which will carry out a wide, deep near-IR spectroscopic survey of galaxies and will target $\sim$50,000 quasars at high redshifts up to $z\sim7$ \citealt{taka14pfs}. Together, these observatories will dramatically expand the known samples of AGN over a wide range in redshift and luminosity, and will enable statistical studies of AGN, their host galaxies, and their large scale structures that will further enhance our understanding of the growth of supermassive black holes over cosmic time.

\begin{acknowledgements}
R.C.H.\ acknowledges support from an Alfred P. Sloan Research Fellowship, a Dartmouth Class of 1962 Faculty Fellowship, the National Science Foundation via grant numbers 1211096 and 1515364, and NASA through ADAP award NNX12AE38G. We are grateful to the organizers for an enjoyable and stimulating Focus Meeting and the IAU for organizing an excellent General Assembly.
\end{acknowledgements}

\end{document}